\documentclass[aps,pra,preprint,groupedaddress]{revtex4-1}

\usepackage{graphicx}
\usepackage{amsmath,amssymb}

\begin{document}

\preprint{KUNS-2407}

\title{When is Multimetric Gravity Ghost-free?}
\author{Kouichi Nomura and Jiro Soda}
\affiliation{Department of Physics, Kyoto University, Kyoto, 606-8502, Japan}

\date{July 16, 2012}

\begin{abstract}
 We study  ghosts in multimetric gravity by combining the mini-superspace and the Hamiltonian constraint analysis. 
We first revisit bimetric gravity and explain why it is ghost-free. 
Then, we apply our method to trimetric gravity and clarify when the model contains a ghost. More precisely,
we prove trimetric gravity generically contains a ghost. However, if we cut the interaction of a pair of metrics, 
trimetric gravity becomes ghost-free. We further extend the Hamiltonian analysis to general multimetric gravity and
calculate the number of ghosts in various models. Thus, we find multimetric gravity with loop type interactions
never becomes ghost-free.
\end{abstract}


\maketitle

\section{Introduction}

It is interesting to explore the possibility that a graviton is massive both from theoretical and phenomenological point of view. 
Theoretically, it is challenging because of various difficulty in constructing a consistent model for a massive graviton.
At the linear level, Fiertz and Pauli succeeded in constructing a ghost-free model for a massive graviton~\cite{Fierz1939}. 
However, it is soon recognized that there is a tension between the theory and experiments, the so-called van Dam-Veltman-Zaharov (vDVZ) discontinuity~\cite{vanDam:1970vg,Zakharov:1970cc}.
It is suggested that the non-linearity resolves the vDVZ discontinuity~\cite{Vainshtein:1972sx}. 
 Unfortunately, it turned out that the non-linearity gives rise to a ghost, the so called Boulware-Deser (BD) ghost~\cite{Boulware1972}. 
Recently,  de Rham, Gabadadze and Tolly have succeeded in constructing ghost-free non-linear massiv gravity theory~\cite{deRham2010a, deRham2010b}
 (See a review \cite{Hinterbichler2011} and references therein.). Still, there remains various theoretically intriguing issues to be explored.
Phenomenologically, there is a chance to explain the current accelerating universe based on massive gravity.
In fact, there appears an effective cosmological constant proportional to the square of graviton mass~\cite{Koyama:2011xz,Koyama:2011yg,Nieuwenhuizen:2011sq,deRham:2010tw,D'Amico:2011jj,Gumrukcuoglu:2011ew}. 
It is worth studying this possibility in detail. 

One peculiar feature of massive theory of graviton is the necessity of a reference metric which breaks the diffeomorphism invariance.
It is natural to promote this reference metric to a dynamical variable, which is nothing but bimetric gravity.
Bimetric gravity contains two metrics $g$ and $f$ interacting each other.
A history of bimetric gravity is long~\cite{Isham:1971gm,Damour:2002wu,Damour:2002ws}. Curiously, bimetric gravity also
suffers from the ghost problem. Thanks to the recent development in massive gravity, however,
 Hassan and Rosen have proposed ghost-free bimetric gravity~\cite{Hassan2011b, Hassan2011c, Hassan2011d}. 
Then, a natural question arises whether or not we can construct ghost-free multimetric gravity. 
Actually, a naive extension of bimetric gravity to trimetric case was proposed in \cite{Nima2012}. 
 There, three metrics $g$, $f$ and $h$  have a  pair interaction between $(g,h)$ ,$(h,f)$ and $(f,g)$, 
which forms a loop structure.  In contrast to bimetric gravity, however, the presence or the absence of BD-ghost remains unknown. 
Recently, Hinterbichler and Rosen showed that  
a large class of multivielbein gravity is ghost free \cite{Hinterbichler2012}. The relation to metric theory is also discussed~\cite{Hassan2012}. 
However, the relation to the models presented in \cite{Nima2012} is not clear. 
The difference between \cite{Nima2012} and \cite{Hinterbichler2012} comes from the loop type interaction. In fact,
the proof by vielbein method is not applicable to the loop type interaction. Hence, we need to study multimetric gravity
with a different approach.

In this paper, we propose a simple method to study the ghost problem and clarify when multimetric gravity is ghost-free.
A method often used for the ghost analysis is to examine models in the decoupling limit. 
However, a more honest way for probing ghosts is to use the Hamiltonian constraint
analysis using the ADM formalism~\cite{Arnowitt:1962hi}.
The difficulty in studying  multimetric gravity with the constraint analysis comes from the existence of a shift vector.
To avoid the difficulty, we employ the mini-superspace approximation.  The mini-superspace reduction of phase space
makes the analysis so simple. Nevertheless, it is sufficient to identify ghosts because this reduction process does not fail
to capture ghosts. 

The organazation of the paper is as follows. 
In section \ref{bimet}, we revisit bimetric gravity and explain our strategy for the ghost analysis.
In section \ref{trimet}, we investigate trimetric gravity using our method and found that the loop type interaction allows ghost.  
In section \ref{mtt}, we further extend the analysis to general $\mathcal{N}$-metric gravity. We clarify when ghost appears in the spectrum. 
The final section \ref{conclusion} is devoted to the conclusion.

\section{BIMETRIC GRAVITY Revisited\label{bimet}}

In this section, we revisit bigravity and explain our method to probe a ghost. 
It is already known that bimetric gravity is ghost-free~\cite{Hassan2011c}. Here, we show the same conclusion can be obtained using
a simple mini-superspace approximation. In the context of massive gravity, the decoupling limit analysis turns out to be
a useful way for the ghost analysis. However, the most complete one is to use Hamiltonian constraint analysis and count physical degrees of freedom.
Our strategy is to use Hamiltonian constraint analysis in the mini-superspace. 

\begin{figure}
\includegraphics[width=4cm]{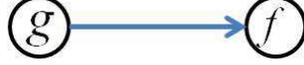}%
\caption{The interaction $\sqrt{g^{-1}f}$ can be represented by a simple diagram. Each blob describes a spacetime with a given metric.
The arrow indicates the order of product. }
\label{fig1}
\end{figure}

 The action of ghost-free bimetric gravity \cite{Hassan2011d} is given by
\begin{eqnarray}
S_{bi}&=&M^2_{g}\int d^4 x\sqrt{-\det g}R[g]+M^2_{f}\int d^4 x\sqrt{-\det f}R[f]  \nonumber\\
     &&  \qquad +2m^2 M_{gf}^2\int d^4 x \sqrt{-\det g}\sum_{n=0}^{4}\beta_n e_n\big(\sqrt{g^{-1}f}\big),
\label{bimetric-action}
\end{eqnarray}
where the first and the second terms are Einstein-Hilbert action for each metric $g$ and $f$ from which we can calculate the scalar curvatures
 $R[g]$, $R[f]$. Here, we have two Planck masses $M_g$ and $M_f$. The last term describes the interaction between two metrics and $\beta_n$
are dimensionless coupling constants. The other constants $m$ and $M_{gf}$ are introduced to adjust the mass dimension.
The square root of the matrix is defined such that $\sqrt{g^{-1}f}\sqrt{g^{-1}f} = g^{\mu\lambda} f_{\lambda\nu}$.
  The interaction terms are constructed by $e_n(\mathbb{X})$ which we define, for matrix $\mathbb{X}$,  
\begin{align}
&e_0(\mathbb{X})=1 \notag \\
&e_1(\mathbb{X})=\mathrm{tr}\mathbb{X} \notag \\
&e_2(\mathbb{X})=\frac{1}{2}\big(\mathrm{tr}^2\mathbb{X}-\mathrm{tr}\mathbb{X}^2\big) \\
&e_3(\mathbb{X})=\frac{1}{6}\big(\mathrm{tr}^3\mathbb{X}-3\mathrm{tr}\mathbb{X}\, \mathrm{tr}\mathbb{X}^2+2\mathrm{tr}\mathbb{X}^3\big) \notag \\
&e_4(\mathbb{X})=\frac{1}{24}\big(\mathrm{tr}^4\mathbb{X}-6\mathrm{tr}^2\mathbb{X}\, \mathrm{tr}\mathbb{X}^2+3\mathrm{tr}^2\mathbb{X}^2+8\mathrm{tr}\mathbb{X}\, \mathrm{tr}\mathbb{X}^3-6\mathrm{tr}\mathbb{X}^4\big)=\det \mathbb{X} 
\ , \notag
\end{align}
where we used the notation $\mathrm{tr}^n \mathbb{X}=\big(\mathrm{tr}\mathbb{X}\big)^n$ and $\mathrm{tr}\mathbb{X}^n=\mathrm{tr}(\mathbb{X}^n)$.
It is useful to represent the interaction by a diagram in Fig.~\ref{fig1}. Note that there is the order between $g$ and $f$ which
is denoted by the arrow.
It is known that the interaction produces massless and massive gravitons and the spectrum is free of Boulware-Deser ghost.
This feature comes from a specific interaction form found in massive gravity theory. Remarkably, there exists the diagonal diffeomorphism invariance
which indicates the presence of the massless graviton. 

Now, let us perform Hamiltonian constraint analysis based on the ADM formalism. In particular,
to make the analysis tractable, we employ the mini-superspace approach.
Namely, we assume spatial homogeneity and express metrics in terms of ADM variables as
\begin{eqnarray}
g_{\mu \nu}dx^{\mu} dx^{\nu}=-N(t)^2dt^2+\gamma_{ij}(t)dx^i dx^j \ ,
\end{eqnarray}
where $N$ is a lapse function and $\gamma_{ij}$ is a spatial metric.
Similarly, we can take the following ansatz
\begin{eqnarray}
f_{\mu \nu}dx^{\mu} dx^{\nu}=-L(t)^2dt^2+\omega_{ij}(t)dx^i dx^j \ ,
\end{eqnarray}
where $L$ is a lapse function and $\omega_{ij}$ is a spatial metric.
It is convenient to write them  in a matrix form,
\begin{align}
g_{\mu \nu}=\left(
\begin{array}{cc}
-N^2 &  0\\
0 & \gamma_{ij}
\end{array}
\right) \ ,\quad 
g^{\mu \nu}=\left(
\begin{array}{cc}
-1/N^2 & 0 \\
 0& \gamma^{ij}
\end{array}
\right) \ ,\notag \\
f_{\mu \nu}=\left(
\begin{array}{cc}
-L^2 &  0\\
0 & \omega_{ij}
\end{array}
\right) \ ,\quad 
f^{\mu \nu}=\left(
\begin{array}{cc}
-1/L^2 & 0 \\
 0& \omega^{ij}
\end{array}
\right) \ ,
\end{align}
where $\gamma^{ij}$ and $\omega^{ij}$  are inverse matrices of spatial metrics $\gamma_{ij}$ and $\omega_{ij}$ . 
Then, a basic part of interaction terms can be calculated to be
\begin{align}
\big(g^{-1}f\big)^{\mu}_{\nu}=
\left(
\begin{array}{cc}
L^2/N^2 & 0 \\
 0& \gamma^{il}\omega_{lj}
\end{array}
\right),\quad 
\sqrt{g^{-1}f}=
\left(
\begin{array}{cc}
L/N & 0 \\
 0& \sqrt{\gamma^{-1}\omega}
\end{array}
\right).
\end{align}

When we count physical degrees of freedom, the following must be taken into account.
In the vacuum cases, we can diagonalize one of two spatial metrics using diagonal spatial coordinate transformations. 
Performing a spatial coordinate transformation  $x^i\rightarrow {\Lambda (t_0)^i}_j\,  x^j$,
we can set one spatial metric at the time $t=t_0$, $\gamma_{ij}(t_0)$, a unit matrix $\delta_{ij}$. 
Moreover, since the orthogonal transformation dose not change $\gamma_{ij}(t_0)=\delta_{ij}$, 
 we can diagonalize $\dot{\gamma}_{ij} (t_0)$ simultaneously by using this freedom. 
At this stage, homogeneous spatial coordinates are completely fixed. 
Now, $\gamma_{ij}$ and $\dot{\gamma}_{ij}$ is diagonal at the time $t=t_0$ as an initial condition. 
Then we assume diagonal form of $\gamma_{ij}(t)$ at all time, and insert it into equations obtained from variations of action. Any contradiction 
 never occurs in vacuum. Thus, we conclude that one spacial metric $\gamma_{ij}(t)$ can be diagonalized because of the uniqueness of the solution. 
Hence, the number of component of one of two metrics reduces from 6 to 3. This fact will be used later. 

In this paper, for simplicity, we assume  that interactions are minimal~\cite{Hassan2011a, Nima2012}, namely
 \begin{align}
 \beta_0=3,\quad   \beta_1=-1,\quad   \beta_2=0,\quad  \beta_3=0, \quad \beta_4=1   \ .
 \end{align}
Clearly, this simplification does not lose any generality concerning with the ghost analysis.
Then, the Lagrangian reads
\begin{eqnarray}
\mathcal{L} =M_g^2\pi^{ij}\dot{\gamma}_{ij}+M_f^2p^{ij}\dot{\omega}_{ij} 
           - NC_N - LC_L \ ,
\end{eqnarray}
where $\pi^{ij}$, $p^{ij}$ are canonical conjugate momentum of $\gamma_{ij}$, $\omega_{ij}$ . Here, we have defined 
\begin{eqnarray}
C_N&=& \frac{M_g^2}{\sqrt{\det \gamma}}\Big(\pi^{ij}\pi_{ij}-\frac{1}{2}{\pi^i}_i{\pi^j}_j\Big)-M_g^2\sqrt{\det \gamma}\, {}^{(3)}\!R[\gamma]     
           +a_1\sqrt{\det \gamma}\big(\mathrm{tr}\sqrt{\gamma^{-1}\omega}-3\big) \\ \notag \\
C_L&=& \frac{M_f^2}{\sqrt{\det \omega}}\Big(p^{ij}p_{ij}-\frac{1}{2}{p^i}_i{p^j}_j\Big)-M_f^2\sqrt{\det \omega}\, {}^{(3)}\!R[\omega]
            +a_1\big(\sqrt{\det \gamma}-\sqrt{\det \omega}\big) \ ,
\end{eqnarray}
where the first two terms of constraints $C_N $ and $C_M$ come from Einstein-Hilbert term in the action, so ${}^{(3)}\!R[\gamma]$ and ${}^{(3)}\!R[\omega]$  are spatial scalar curvatures computed from $\gamma$ and $\omega$, respectively. 
The last term of each constraint comes from the interaction (see the derivation in Appendix A), and we use $a_1=2m^2M_{gf}^2$. Now, the Hamiltonian is given by
\begin{align}
H = NC_N + LC_L   \ .
\end{align}
 Since there are two Lagrange multipliers, there are two primary constraints 
\begin{align}
C_N=0,\qquad C_L=0  \ .
\end{align}
Moreover, we need to impose consistency conditions for them 
\begin{align}
\dot{C}_N&=\big\{C_N\, ,\, H\big\}=L\big\{C_N\, ,\, C_L\big\} \equiv LC_{NL}\approx0 \ , \notag \\
\dot{C}_L&=\big\{C_L\, ,\, H\big\}=N\big\{C_L\, ,\, C_N\big\} \equiv NC_{LN}\approx0 \ ,
\label{tricon} 
\end{align}
 where the Poisson bracket $\{F,G\}$ is  defined by
\begin{align}
\big\{F\, ,\, G\big\}=& \Big(\frac{\partial F}{\partial \gamma_{mn}}\frac{\partial G}{\partial \pi^{mn}}-\frac{\partial F}{\partial \pi^{mn}}\frac{\partial G}{\partial \gamma_{mn}}\Big)
           +\Big(\frac{\partial F}{\partial \omega_{mn}}\frac{\partial G}{\partial p^{mn}}-\frac{\partial F}{\partial p^{mn}}\frac{\partial G}{\partial \omega_{mn}}\Big).
\end{align}
Here, "$\approx 0$" means "$=0$" on the constraint surface. Notice that $\{F,F\}=0$ because of spatial homogeneity.

 To check if secondary constraint arises or not, we have to calculate Poisson bracket $C_{NL}$. 
From the calculation presented in Appendix B, we obtain 
\begin{align}
C_{NL}=\big\{C_N\, ,\, C_L\big\}=a_1\bigg[\frac{1}{2}M_g^2{\pi^i}_i-M_f^2\sqrt{\frac{\det \gamma}{\det \omega}}\, \Big(\frac{1}{2}{p^i}_i\, \mathrm{tr}\sqrt{\gamma^{-1}\omega}-\mathrm{tr}\big(\sqrt{\gamma^{-1}\omega}\, p\, \omega\big)\Big)\bigg].
\end{align}
This leads to one secondary constraint $C_{NL}\approx 0$.  The consistency condition for the secondary constraint
reads 
\begin{align}
\dot{C}_{NL}=N\big\{C_{NL}\, ,\, C_N\big\} + L\big\{C_{NL}\, ,\, C_J\big\}\approx 0 \ .
\end{align}
This condition determines one of two Lagrange multipliers $N$ and $L$. The remaining multiplier describes the diagonal 
time reparametrization invariance in bimetric gravity.

The number of components of two metrics and their canonical conjugates is $24$. Since we can diagonalize one of the two metrics, 
we should subtract 6 from this number. Recall that there are two primary constraints and one secondary constraint. 
Furthermore, as we have one first class constraint, we have to put one gauge condition.
Thus, the total number of degrees of freedom should be $(24-6-2-1-1)/2=7$ in configuration space, 
which matches degrees of freedom of one massless graviton and one massive graviton. 
This proves that BD ghost is absent in bimetric gravity described by the action (\ref{bimetric-action}).

\section{TRIMETRIC GRAVITY\label{trimet}}

Now, we apply the method explained in the previous section to trimetric gravity.
In contrast to the bimetric gravity, there are two kind of interactions, namely, the tree type and the loop type interactions.
We discuss both cases, separately. 

The action for trimetric gravity~\cite{Nima2012} can be written as
\begin{eqnarray}
S_{tri}&=&M^2_{g}\int d^4 x\sqrt{-\det g}R[g]+M^2_{f}\int d^4 x\sqrt{-\det f}R[f]+M^2_{h}\int d^4 x\sqrt{-\det h}R[h] \notag \\
        && \qquad +2m_1^2 M_{gf}^2\int d^4 x\sqrt{-\det g}\sum_{n=0}^{4}\beta_n e_n\big(\sqrt{g^{-1}f}\big) \notag \\
        && \qquad +2m_2^2 M_{fh}^2\int d^4 x \sqrt{-\det f} \sum_{n=0}^{4}\beta'_n e_n\big(\sqrt{f^{-1}h}\big) \notag \\
        && \qquad +2m_3^2 M_{hg}^2\int d^4 x \sqrt{-\det h} \sum_{n=0}^{4}\beta''_n e_n\big(\sqrt{h^{-1}g}\big),
\end{eqnarray}
where $\beta_n$, $\beta_n'$ and $\beta_n''$ are free parameters and $R[g]$, $R[f]$ and $R[h]$ are scalar curvatures
constructed from metrics $g$, $f$ and $h$, respectively. 
We also introduced new mass parameters $m_1 , m_2 , m_3 , M_{fh} , M_{hg} $ and a Planck mass $M_h$. 
It should be noted that there exists the diagonal diffeomorphism invariance in this trimetric theory which makes one of gravitons massless. 
As discussed in \cite{Nima2012, Hinterbichler2012}, if this trimetric gravity contains no extra degrees of freedom, the total number of degrees of freedom should be $2+5+5=12$, which comes from one massless graviton and two massive gravitons. 
 From now on, we use
\begin{align}
a_1=2m_1^2 M_{gf}^2\, ,\quad   a_2=2m_2^2 M_{fh}^2\, ,\quad  a_3=2m_3^2 M_{hg}^2
\end{align}
for notational simplicity. If we have $a_1\neq0$, $a_2\neq0$ and $a_3\neq0$, all pairs $(g,f)$, $(f,h)$ and $(h,g)$ interact and we call it the loop type interaction. When one of $a_i \: (i=1,2,3)$ is set to zero, two of three pairs of interactions remain, which we call the tree type interaction. 
The case where one interaction is cut is already proved to be ghost-free using vielbein formalism~\cite{Hinterbichler2012},
however, for the loop type interaction no one shows the presence or the absence of ghost. 
In this paper, we settle this issue.

 Apparently, the full Hamiltonian constraint analysis is difficult.
To circumvent this difficulty, we take the method used in the previous section.
Namely, we assume spatial homogeneity and express metrics in terms of ADM variables as
\begin{eqnarray}
g_{\mu \nu}dx^{\mu} dx^{\nu}=-N(t)^2dt^2+\gamma_{ij}(t)dx^i dx^j \ ,
\end{eqnarray}
where $N$ is a lapse function and $\gamma_{ij}$ is a spatial metric.
Similarly, we can take the following ansatz
\begin{eqnarray}
f_{\mu \nu}dx^{\mu} dx^{\nu}=-L(t)^2dt^2+\omega_{ij}(t)dx^i dx^j \ ,
\end{eqnarray}
and
\begin{eqnarray}
h_{\mu \nu}dx^{\mu} dx^{\nu}=-Q(t)^2dt^2+\rho_{ij}(t)dx^i dx^j \ ,
\end{eqnarray}
where $L$ and $Q$ are lapse functions and $\omega_{ij}$ and $ \rho_{ij}$ are spatial metrics.
 To perform Hamiltonian constraint analysis, we need the Lagrangian in the ADM variables
\begin{eqnarray}
\mathcal{L} =M_g^2\pi^{ij}\dot{\gamma}_{ij}+M_f^2p^{ij}\dot{\omega}_{ij}+M_h^2\phi^{ij}\dot{\rho}_{ij} 
           - NC_N - LC_L - QC_Q \ ,
\end{eqnarray}
where $\pi^{ij}$, $p^{ij}$ and $\phi^{ij}$ are canonical conjugate momentum of $\gamma_{ij}$, $\omega_{ij}$ and $\rho_{ij}$. 
Here, three Hamiltonian constraints 
\begin{eqnarray}
C_N&=& \frac{M_g^2}{\sqrt{\det \gamma}}\Big(\pi^{ij}\pi_{ij}-\frac{1}{2}{\pi^i}_i{\pi^j}_j\Big)-M_g^2\sqrt{\det \gamma}\, {}^{(3)}\!R[\gamma] \notag \\
    &&+a_1\sqrt{\det \gamma}\big(\mathrm{tr}\sqrt{\gamma^{-1}\omega}-3\big)+a_3\big(\sqrt{\det \rho}-\sqrt{\det \gamma}\big) \ ,\\ \notag \\
C_L&=& \frac{M_f^2}{\sqrt{\det \omega}}\Big(p^{ij}p_{ij}-\frac{1}{2}{p^i}_i{p^j}_j\Big)-M_f^2\sqrt{\det \omega}\, {}^{(3)}\!R[\omega] \notag \\
    &&+a_2\sqrt{\det \omega}\big(\mathrm{tr}\sqrt{\omega^{-1}\rho}-3\big)+a_1\big(\sqrt{\det \gamma}-\sqrt{\det \omega}\big) \ , 
\end{eqnarray}
and
\begin{eqnarray}
C_Q&=& \frac{M_h^2}{\sqrt{\det \rho}}\Big(\phi^{ij}\phi_{ij}-\frac{1}{2}{\phi^i}_i{\phi^j}_j\Big)-M_h^2\sqrt{\det \rho}\, {}^{(3)}\!R[\rho] \notag \\
    &&+a_3\sqrt{\det \rho}\big(\mathrm{tr}\sqrt{\rho^{-1}\gamma}-3\big)+a_2\big(\sqrt{\det \omega}-\sqrt{\det \rho}\big)
\end{eqnarray}
emerge. The first line of each Hamiltonian constraint comes from the
Einstein-Hilbert term in the action, so ${}^{(3)}\!R[\gamma]$, ${}^{(3)}\!R[\omega]$ and ${}^{(3)}\!R[\rho]$ are spatial scalar curvatures 
calculated from $\gamma$, $\omega$ and $\rho$, respectively. The other terms can be derived as explained in Appendix A.
Then, the Hamiltonian can be read off as
\begin{align}
H = NC_N + LC_L + QC_Q \ .
\end{align}
 Since there are three Lagrange multipliers, there arise three primary constraints 
\begin{align}
C_N=0,\qquad C_L=0,\qquad C_Q=0  \ .
\label{primary-NLQ}
\end{align}
Moreover, we need consistency conditions for them 
\begin{align}
\dot{C}_N&=\big\{C_N\, ,\, H\big\}=L\big\{C_N\, ,\, C_L\big\}+Q\big\{C_N\, ,\,C_Q\big\} \approx0 \ ,\notag \\
\dot{C}_L&=\big\{C_L\, ,\, H\big\}=N\big\{C_L\, ,\, C_N\big\}+Q\big\{C_L\, ,\,C_Q\big\} \approx0 \ ,\label{tricon} \\
\dot{C}_Q&=\big\{C_Q\, ,\, H\big\}=N\big\{C_Q\, ,\, C_N\big\}+L\big\{C_Q\, .\,C_L\big\} \approx0 \ .\notag 
\end{align}
 To check if secondary constraints arise, we must calculate Poisson brackets. 
From the calculation in Appendix B, we obtain
\begin{align}
C_{NL}\equiv 
\big\{C_N\, ,\, C_L\big\}=a_1\bigg[\frac{1}{2}M_g^2{\pi^i}_i-M_f^2\sqrt{\frac{\det \gamma}{\det \omega}}\, \Big(\frac{1}{2}{p^i}_i\, \mathrm{tr}\sqrt{\gamma^{-1}\omega}-\mathrm{tr}\big(\sqrt{\gamma^{-1}\omega}\, p\, \omega\big)\Big)\bigg].
\end{align}
By performing permutations among $g=(N,\gamma)\: $, $f=(L,\omega)$ and $h=(Q,\rho)$, we also get
\begin{eqnarray}
C_{LQ}\equiv
\big\{C_L\, ,\, C_Q\big\}=a_2\bigg[\frac{1}{2}M_f^2{p^i}_i-M_h^2\sqrt{\frac{\det \omega}{\det \rho}}\, \Big(\frac{1}{2}{\phi^i}_i\, \mathrm{tr}\sqrt{\omega^{-1}\rho}-\mathrm{tr}\big(\sqrt{\omega^{-1}\rho}\, \phi\, \rho\big)\Big)\bigg] 
\end{eqnarray}
and
\begin{eqnarray}
C_{QN}\equiv
\big\{C_Q\, ,\, C_N\big\}=a_3\bigg[\frac{1}{2}M_h^2{\phi^i}_i-M_g^2\sqrt{\frac{\det \rho}{\det \gamma}}\, \Big(\frac{1}{2}{\pi^i}_i\, \mathrm{tr}\sqrt{\rho^{-1}\gamma}-\mathrm{tr}\big(\sqrt{\rho^{-1}\gamma}\, \pi\, \gamma\big)\Big)\bigg] \ .
\end{eqnarray}
In general, quantities inside the bracket does not vanish.
Hence, the coefficients $a_1 , a_2 $ and $a_3$ determine the consistency conditions.

\subsection{Tree Type Interaction}

\begin{figure}
\includegraphics[width=4cm]{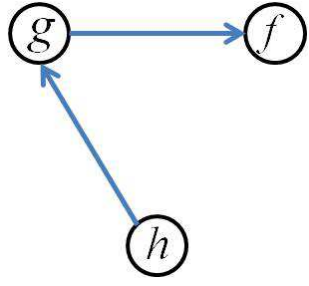}%
\caption{The diagram represents the tree type interaction.}
\label{fig2}
\end{figure}

 In this subsection, we consider the tree type interaction 
\begin{align}
a_1 \neq 0, \quad a_2=0,\quad  a_3 \neq 0 \ ,
\end{align}
which cut interaction between $f$ and $h$ as in Fig.\ref{fig2}. 
In any case, there are primary constraints (\ref{primary-NLQ}). Since $C_{QL}=C_{LQ}=0$
trivially holds,  consistency conditions (\ref{tricon}) lead to equations
\begin{align}
LC_{NL}+QC_{NQ} \approx 0,\quad NC_{LN}\approx 0,\quad NC_{QN}\approx 0 \ .
\end{align}
Hence, we have two secondary constraints 
\begin{align}
C_{NL}\approx 0,\quad C_{NQ}\approx 0 \ .
\end{align}
Moreover, we must impose consistency conditions
\begin{align}
\dot{C}_{NL}&=\big\{C_{NL}\, ,\, H \big\}=N\big\{C_{NL}\, ,\, C_N \big\}+L\big\{C_{NL}\, ,\, C_L \big\}+Q\big\{C_{NL}\, ,\, C_Q \big\}\approx 0 \notag \\
\dot{C}_{QL}&=\big\{C_{QL}\, ,\, H \big\}=N\big\{C_{QL}\, ,\, C_N \big\}+L\big\{C_{QL}\, ,\, C_L \big\}+Q\big\{C_{QL}\, ,\, C_Q \big\}\approx 0,
\end{align}
which determine two of three Lagrange multipliers $N$, $L$ and $Q$. The remaining multiplier is related  to the gauge transformation. 

Eventually, we have five constraints and one gauge freedom.  In trimetric gravity, propagating modes are spatial metrics. 
Each of them has six components, but as is already explained
we can diagonalize one of them. Hence, trimetric gravity has $3+6+6=15$ degrees of freedom in configuration space and $15\times 2=30$ in phase apace.
 Thus, the total number of degrees of freedom is $(30-5-1)/2=12$ which matches the physical degrees of
one massless and two massive gravitons. Therefore, no BD ghost exists in the spectrum. This conclusion 
is consistent with the one obtained by the vielbein method~\cite{Hinterbichler2012}.

\subsection{Loop Type Interaction}

\begin{figure}
\includegraphics[width=4cm]{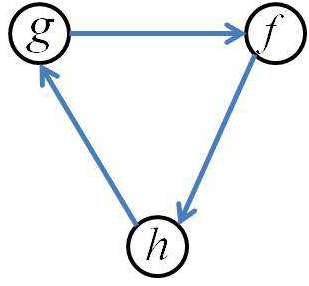}%
\caption{The diagram represents the loop type interaction.}
\label{fig3}
\end{figure}

Now, we consider the more general loop type interaction represented by a diagram in Fig.\ref{fig3}
\begin{align}
a_1\neq 0,\quad a_2\neq 0,\quad a_3\neq 0 \ .
\end{align}
It is obvious that
\begin{align}
\big\{C_N\, ,\, C_L\big\}\neq 0,\quad \big\{C_L\, ,\, C_Q\big\}\neq 0,\quad \big\{C_Q\, ,\, C_N\big\}\neq 0
\end{align}
even on the constraint surface. Hence,  consistency conditions (\ref{tricon}) do not generate any secondary constraint.
Instead, it determines Lagrange multipliers $N$, $L$ and $Q$.
However, due to the antisymmetric property of Poisson brackets
\begin{align}
C_{NL}&=\big\{C_N\, ,\, C_L\big\}=-\big\{C_L\, ,\, C_N\big\}=C_{LN} \\
C_{LQ}&=\big\{C_L\, ,\, C_Q\big\}=-C_{QL}\\
C_{QN}&=\big\{C_Q\, ,\, C_N\big\}=-C_{NQ},
\end{align}
only two of them are determined. For example, choosing 
\begin{align}
L=-\frac{C_{NQ}}{C_{NL}}\, Q,\quad N=-\frac{C_{LQ}}{C_{LN}}\, Q,
\end{align}
all of consistency conditions (\ref{tricon}) are satisfied. 

To conclude, we have three primary constraints and we need one gauge condition to fix
one undetermined Lagrange multiplier which is associated with the time reparametrization invariance. 
 In trimetric gravity, as is already counted, there are $3+6+6=15$ degrees of freedom in configuration space and $15\times 2=30$ in phase apace. 
In phase space, we have three constraints and one gauge condition, so total number of degrees of freedom is $(30-3-1)/2=13$. 
If no BD ghost is present, there must be
$2+5+5=12$ degrees of freedom which comes from one massless graviton and two massive gravitons. Therefore, one extra degree of freedom exists
 and it should be a BD ghost.
Thus, we have proved the existence of a ghost in generic trimetric gravity.

\section{General Multimetric Models \label{mtt}}

Now, we are in a position to discuss more general cases.
We explicitly calculate the number of ghosts if they exist.

In this section, we consider $\mathcal{N}$ dynamical metrics $g_k$  ($k=1,2,..,\mathcal{N}$) and interaction terms such as
\begin{align}
\sum_{k=1}^{\mathcal{N}}a_k\sqrt{-\det g_k}\sum_{n=0}^4\beta_{k,n}e_n\Big(\sqrt{g_k^{-1}g_{k+1}}\Big),
\end{align}
where we define $g_{\mathcal{N}+1}=g_1$ and for later purpose we also need  $g_{0}=g_{\mathcal{N}}$. 
Let us describe the interaction between two metrics $g_k$ and $g_{k+1}$ in terms of ADM form of metrics 
\begin{eqnarray}
  ds_k^2 = -N_k^2 (t) dt^2 + \gamma_{k, ij} (t) dx^i dx^j \ .
\end{eqnarray}
Schematically, the interaction can be written as
\begin{align}
\sqrt{-\det g_k}\sum_{n=0}^4\beta_{k,n}e_n\Big(\sqrt{g_k^{-1}g_{k+1}}\Big)=N_{k}F_k\big(\gamma_k : \gamma_{k+1}\big)+N_{k+1}G_k\big(\gamma_k : \gamma_{k+1}\big),
\end{align}
where $F_k$ and $G_k$ are some functions determined by parameters $\beta_{k,n}$. Thus, the total interaction terms are given by
\begin{align}
\sum_{k=1}^{\mathcal{N}}a_k\sqrt{-\det g_k}\sum_{n=0}^4\beta_{k,n}e_n\Big(\sqrt{g_k^{-1}g_{k+1}}\Big)
=&\sum_{k=1}^{\mathcal{N}}N_k\Big\{a_kF_k\big(\gamma_k : \gamma_{k+1}\big)+a_{k-1}G_{k-1}\big(\gamma_{k-1} : \gamma_k\big)\Big\} \ .
\end{align}
The Hamiltonian becomes
\begin{align}
H=\sum_{k=1}^{\mathcal{N}}N_k C_k,\qquad C_k=C^0_k\big(\gamma_k , \pi_k\big)-a_kF_k\big(\gamma_k : \gamma_{k+1}\big)-a_{k-1}G_{k-1}\big(\gamma_{k-1} : \gamma_k\big)  \ ,
\end{align}
where $C^0_k$ comes from the Einstein Hilbert term for $g_k$, so it contains $\gamma_k$ and its canonical cojugate momentum $\pi_k$. 

Corresponding to $\mathcal{N}$ Lagrange multipliers, we have $\mathcal{N}$ primary constraints 
\begin{eqnarray}
C_k=0  \ , \qquad  (k=1,2,..,\mathcal{N}) \ .
\end{eqnarray}
 Next, we have to examine $\mathcal{N}$ consistency conditions   
\begin{align}
\dot{C_k}=C_{k,k-1}N_{k-1}+C_{k,k+1}N_{k+1}\approx 0, \label{con}
\end{align}
where $C_{k,l}=\big\{C_k\, , \, C_l\big\}$  and $C_{k,l}=0$ if $|k-l|\geqq 2$. 
In this formula, $N_0 = N_\mathcal{N}$ and $N_{\mathcal{N}+1} = N_1$ should be understood.
Note that the explicit calculation gives rise to an important information
\begin{eqnarray}
C_{k,k+1}\propto a_k  \ .
\end{eqnarray}
The structure of this matrix depends on odd or even number. 
For example, in the case $\mathcal{N}=4$, we have
\begin{align}
C_{k,l}=\left(
\begin{array}{cccc}
0 & C_{1,2}& 0&  C_{1,4}\\
-C_{1,2} & 0&C_{2,3} &0  \\
 0& -C_{2,3}& 0& C_{3,4}\\
 -C_{1,4}& 0& -C_{3,4} & 0
\end{array}
\right). \label{mat4}
\end{align}
While, in the case of $\mathcal{N}=5$, we get
\begin{align}
C_{k,l}=\left(
\begin{array}{ccccc}
0 & C_{1,2}& 0& 0&  C_{1,5}\\
-C_{1,2} & 0&C_{2,3} &0 & 0 \\
 0& -C_{2,3}& 0& C_{3,4}& 0\\
 0& 0& -C_{3,4}& 0& C_{4,5}\\
 -C_{1,5}& 0& 0& -C_{4,5}& 0
\end{array}
\right). \label{mat5}
\end{align}
In the case of odd number of metrics, we cannot split the equations into two independent sets.
While, in the case of even number of metrics, we can split a set of equations into independent two groups of equations.
Hence, we have to discuss two cases, separately. 

\subsection{Tree Type Interaction}

\begin{figure}
\includegraphics[width=4cm]{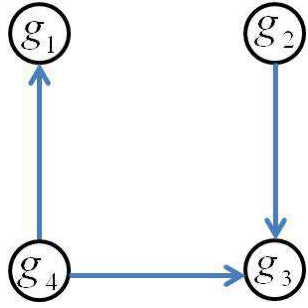}%
\caption{The diagram represents the tree type interaction.}
\label{fig4}
\end{figure}

First, we consider the tree type interaction.

If we cut one of $(g_k,g_{k+1})_{k=1,2,..,\mathcal{N}}$ interactions as in Fig.\ref{fig4}, 
for example setting $a_1=0$, Eq.(\ref{con}) leads to $\mathcal{N}-1$ secondary constraints
\begin{align}
C_{k,k+1} \approx 0,\: (k=2,3,..,\mathcal{N}),
\end{align}
and their consistency conditions
\begin{align}
\dot{C}_{k,k+1}= \sum_{l=1}^{\mathcal{N}}\big\{{C}_{k,k+1}\, ,\, C_l\big\}N_l \approx 0,\: (k=2,3,..,\mathcal{N})
\end{align}
determine $\mathcal{N}-1$ of $N_k$ ($k=1,2,..,\mathcal{N}$), only one Lagrange multiplier remains undetermined. 
Therefore, the total number of degrees of freedom can be deduced as
\begin{align}
\frac{1}{2}\Big(2\big(3+6(\mathcal{N}-1)\big)-\mathcal{N}-(\mathcal{N}-1)-1\Big)=5(\mathcal{N}-1)+2,
\end{align}
which corresponds to $\mathcal{N}-1$ massive and one massless gravitons. Therefore, there exists no BD ghost. 
This conclusion is also consistent with the one obtained  by the vielbein method~\cite{Hinterbichler2012}.

\subsection{Loop Type Interaction}

\begin{figure}
\includegraphics[width=4cm]{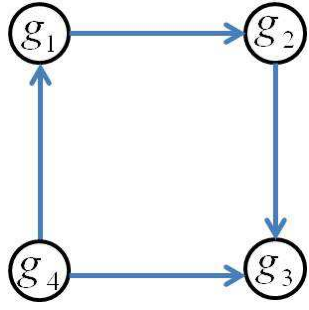}%
\caption{The diagram represents the loop type interaction.}
\label{fig5}
\end{figure}

Now, we come to our main point.

If all of $(g_k,g_{k+1})_{k=1,2,..,\mathcal{N}}$ interactions exist as in Fig.\ref{fig5}, 
the analysis gets a little complicated. We have to discuss odd and even numbers, separately.

\subsubsection{Odd Number of Metrics}

First, we consider the case where $\mathcal{N} = 2m+1$, where $m$ is a natural number. 
In this case, we can classify Eq.(\ref{con}) into the following four parts 
\begin{eqnarray}
&& C_{2k,2k-1}N_{2k-1}+C_{2k,2k+1}N_{2k+1}= 0 \qquad (k=1,2,3,...,m) \label{odd1} \\
&& C_{2k-1,2k-2}N_{2k-2}+C_{2k-1,2k}N_{2k}= 0 \qquad (k=2,3,...,m) \label{odd2} \\
&& C_{1,2m+1}N_{2m+1}+C_{1,2}N_{2}= 0 \label{odd3} \\
&& C_{2m+1,2m}N_{2m}+C_{2m+1,1}N_{1}= 0 \label{odd4} \ .
\end{eqnarray}
Solving Eq.(\ref{odd1}), we see all of $N_{2k+1}$ ($k=1,2,..,m$) can be expressed by $N_{1}$. 
Similarly, Eq.(\ref{odd2}) can be used to express $N_{2k}$ ($k=2,3,..,m$) in terms of $N_2$. 
Substituting these results into Eq.(\ref{odd3}) and Eq.(\ref{odd4}), we obtain a single equation
which determines $N_2 $ by $N_1$. Thus,  
Eq.(\ref{con}) determines $\mathcal{N}-1$ Lagrange multipliers, and one multiplier is left undetermined, which reflects
 the existence of gauge symmetry.

In the case of odd number of metrics, there is no secondary constraint. While, we need one gauge condition to fix the gauge degree of freedom.
 In conclusion, the total number of degrees of freedom can be calculated as
\begin{align}
\frac{1}{2}\Big(2\big(3+6(\mathcal{N}-1)\big)-\mathcal{N}-1\Big)=5(\mathcal{N}-1)+2+\frac{\mathcal{N}-1}{2} \ .
\end{align}
Here, the first two terms correspond to massive and massless gravitons, respectively.
The last one should be BD ghosts
and the number of ghosts is given by $(\mathcal{N}-1)/2$.

\subsubsection{Even Number of Metrics}

Next, we consider the case $\mathcal{N} = 2m+2 $, where $m$ is a natural number.  
In this case, we can split Eq.(\ref{con}) into two independent sets of equations, 
\begin{align}
& C_{k,k-1}N_{k-1}+C_{k,k+1}N_{k+1}=0 \qquad (k=1,3,5,.., 2m+1) \label{eve1}\\
& C_{k,k-1}N_{k-1}+C_{k,k+1}N_{k+1}=0 \qquad (k=2,4,6,.., 2m+2) \label{eve2}.
\end{align}
The first set (\ref{eve1}) contains only $N_k$ $(k=2,4,6,..,2m+2)$, and the second set (\ref{eve2}) contains $N_k$ $(k=1,3,5,..,2m+1)$. 
Here, if the component $C_{k,k\pm1}$ is in Eq.(\ref{eve1}), $C_{k\pm1,k}=-C_{k,k\pm1}$ must be in Eq.(\ref{eve2}) and vice versa. Therfore, 
in each set, every component $C_{k,k\pm1}$ appears only once.
Now, we define 
\begin{align}
&D_{i,j}=C_{2i-1,2j}\, ,\qquad  M_j=N_{2j}\qquad (i,j=1,2,3,..,m+1).
\end{align}
Note that $D_{ij}\neq 0$ only for $i-j=0,1$. Then, Eq.(\ref{eve1}) can be written as 
\begin{align}
\sum_j D_{i,j}M_j=0 \qquad (i=1,2,3,..,m+1), 
\end{align}
which we can split into 
\begin{align}
&D_{1,1}M_1+D_{1,m+1}M_{m+1}=0 \label{eve3} \\
&D_{i,i-1}M_{i-1}+D_{i,i}M_i=0 \qquad (i=2,3,..,m+1) \label{eve4}.
\end{align}
Using Eq.(\ref{eve4}), we can solve all of $M_j$ $(i=2,3,..,m+1)$ in terms of $M_1$. However, the relation between $M_1$ and $M_{m+1}$ 
obtained from Eq.(\ref{eve4}) is not the same as Eq.(\ref{eve3}) because Eq.(\ref{eve4})  contains no $D_{1,1}$ and $D_{1,m+1}$. 
So, we have to impose a constraint so that we get non-trivial Lagrange multipliers. 
This is a secondary constraint expressed by
\begin{align}
\det D_{ij}=0 \ .
\end{align} 
Under this condition, $m$ of $M_j$ $(j=1,2,..,m+1)$ are determined, and one is left undetermined. 

Now, we take latter set (\ref{eve2}) and define  
\begin{align}
&E_{i,j}=C_{2i,2j-1}\, ,\qquad  W_j=N_{2j-1}\qquad (i,j=1,2,3,..,m+1).
\end{align}
The same argument applies, so we get a secondary constraint $\det{E_{ij}}=0$, and one of $W_j$ $(j=1,2,..,m+1)$ is left undetermined. 
However, matrix $E_{ij}$ satisfies $E_{ij}=-D_{ji}$. Hence, $\det E_{ij}=0$ is not a new constraint. 
Therefore, from Eq.(\ref{con}), we get one secondary constraint $\det D_{ij}=0$ and two undetermined Lagrange multipliers . 
Then, we must impose a consistency condition for the secondary constraint
\begin{align}
\frac{d}{dt}\det D_{ij} =\sum_{k=1}^{\mathcal{N}}\big\{ \det D_{ij} \, ,\, C_k\big\}N_k \approx 0,
\end{align}
which reduces the number of undetermined Lagrange multipliers from two to one. 

To summarize, there are $\mathcal{N}$ primary constraints and one secondary constraint and we need one gauge condition.
Thus, we come to the conclusion that the total number of degrees of freedom is
\begin{align}
\frac{1}{2}\Big(2\big(3+6(\mathcal{N}-1)\big)-\mathcal{N}-1-1\Big)=5(\mathcal{N}-1)+2+\frac{\mathcal{N}-2}{2} \ .
\end{align}
Here, again, the first two terms correspond to massive and massless gravitons, respectively.
Hence, the number of BD ghosts should be $(\mathcal{N}-2)/2$.

\subsection{More General Diagrams}

\begin{figure}
\includegraphics[width=10cm]{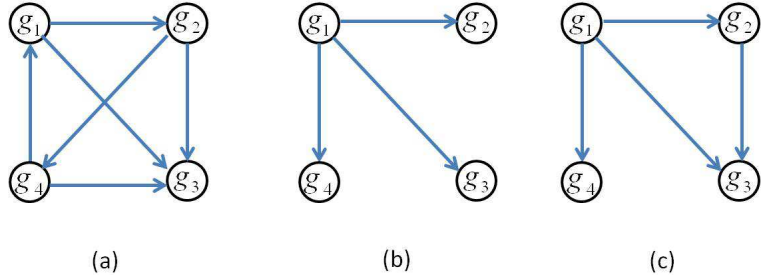}%
\caption{The diagram (a) represents the most general type interaction. The diagram (b) is the broom type
interaction. The diagram (c) includes the loop type interaction, hence there should be a BD ghost.}
\label{fig6}
\end{figure}

In the previous sections, we have considered tree and loop type interactions.
In the case of bimetric gravity, the interaction type is unique, namely, there is only the tree type interaction.
In the case of trimetric gravity, there are two possibilities, the tree and the loop type interaction.
In the case of tetrametric gravity, there are many loop type interaction represented by a diagram (a) in Fig.\ref{fig6}.
If we cut some of the interaction, we can make the tree type interaction and the broom type interaction 
represented by a diagram (b) in Fig.\ref{fig6}.
From our analysis, it is apparent that if the interaction contains at least a loop, then there are ghosts.
For example, the model with a diagram (c) in Fig.\ref{fig6} contains a ghost.
Therefore, in generic cases, there exist ghosts in multimetric gravity. 
The number of ghosts depends on the interaction pattern. 
To construct a viable model,
we have to eliminate all of loop type interactions.

\section{CONCLUSION\label{conclusion}}

We studied  multimetric gravity by combining the mini-superspace and the Hamiltonian constraint analysis. 
We first revisited bimetric gravity and explained why it is ghost-free. 
This proved validity of our method. 
Then, we applied our method to trimetric gravity and clarified when the model contains a ghost. 
We proved trimetric gravity generically contains a ghost. However, if we cut the interaction of a pair of metrics, 
trimetric gravity turned out to be ghost-free. We further extended the Hamiltonian analysis to general multimetric gravity and
calculate the number of ghosts in various models. Thus, we found multimetric gravity with loop type interactions
never becomes ghost-free.
The number of BD ghost in $\mathcal{N}$ metric case turned out to be $(\mathcal{N}-1)/2$ or $(\mathcal{N}-2)/2$, depending on whether
the number of metrics $\mathcal{N}$ is odd or even.
Hence, the number of BD ghosts increases by one everytime two more metrics are introduced. 
There are other models which may contain ghosts or may not contain any ghost.
It depends on the interaction type. The number of ghosts can be calculated once the diagram characterizing the interaction pattern
is given.

Admittedly, what we have investigated is BD ghosts. There may be other ghosts depending on the solutions~\cite{DeFelice:2012mx,Gumrukcuoglu:2012aa}.
In other words, the absence of BD ghosts is a necessary condition as a healthy model.
In this paper, we have studied interaction terms consists of only pairs of metrics. However, as in \cite{Hinterbichler2012}, 
interactions of triplets or quadruplets may be allowed. We hope to study this possibility in future. 
It is also interesting to extend our analysis to higher curvature theories~\cite{Paulos:2012xe}.

\begin{acknowledgments}
This work was supported in part by the Japan Society for the Promotion of Science (JSPS) grant No. 24 -1693,
 the Grant-in-Aid for  Scientific Research Fund of the Ministry of 
Education, Science and Culture of Japan No.22540274, the Grant-in-Aid
for Scientific Research (A) (No.21244033, No.22244030), the
Grant-in-Aid for  Scientific Research on Innovative Area No.21111006,
JSPS under the Japan-Russia Research Cooperative Program,
the Grant-in-Aid for the Global COE Program 
``The Next Generation of Physics, Spun from Universality and Emergence".
\end{acknowledgments}

\appendix

\section{Interaction Terms}

In this appendix, we calculate the interaction terms separately.
We use the following representations
\begin{align}
\big(g^{-1}f\big)^{\mu}_{\nu}=
\left(
\begin{array}{cc}
L^2/N^2 & 0 \\
 0& \gamma^{il}\omega_{lj}
\end{array}
\right),\quad 
\sqrt{g^{-1}f}=
\left(
\begin{array}{cc}
L/N & 0 \\
 0& \sqrt{\gamma^{-1}\omega}
\end{array}
\right).
\end{align}
The first one is given by
\begin{eqnarray}
\sqrt{-\det g}\, e_1\big(\sqrt{g^{-1}f}\big)&=&\sqrt{-\det g}\, \mathrm{tr}\sqrt{g^{-1}f} \notag \\
                                         &=&\sqrt{\det \gamma}\, N\Big(L/N+\mathrm{tr}\sqrt{\gamma^{-1}\omega}\Big) \notag \\
                                         &=&\sqrt{\det \gamma}\, \Big(L+N\mathrm{tr}\sqrt{\gamma^{-1}\omega}\Big)  \ .
\end{eqnarray}
Due to the combination $g^{-1} f$, we got the linear terms with respect to the lapse functions.
The second one becomes
\begin{eqnarray}
\sqrt{-\det g}\, e_2\big(\sqrt{g^{-1}f}\big)&=&\sqrt{-\det g}\, \frac{1}{2}\Big({\mathrm{tr}}^2\sqrt{g^{-1}f}-\mathrm{tr}\big(g^{-1}f\big)\Big) \notag \\
                                         &=&\sqrt{\det \gamma}\, N\, \frac{1}{2}\Big\{\Big(L/N+\mathrm{tr}\sqrt{\gamma^{-1}\omega}\Big)^2-L^2/N^2-\mathrm{tr}\big(\gamma^{-1}\omega \big) \Big\} \notag \\
                                         &=&\sqrt{\det \gamma}\, \Big\{L\, \mathrm{tr}\sqrt{\gamma^{-1}\omega}+\frac{1}{2}N\Big({\mathrm{tr}}^2\sqrt{\gamma^{-1}\omega}-\mathrm{tr}\big(\gamma^{-1}\omega \big)\Big)\Big\} \ .
\end{eqnarray}
Again, we obtained desired linearity for the lapse functions.
The third one can be calculated as
\begin{eqnarray}
\sqrt{-\det g}\, e_3\big(\sqrt{g^{-1}f}\big)&=&\sqrt{-\det g}\, \frac{1}{6}\Big({\mathrm{tr}}^3\sqrt{g^{-1}f}-3\mathrm{tr}\sqrt{g^{-1}f}\, \mathrm{tr}\big(g^{-1}f \big)+2\mathrm{tr}\big(g^{-1}f \big)^{3/2}\Big) \notag \\
                                            &=&\sqrt{\det \gamma}\, N\, \frac{1}{6}\Big\{\Big(L/N+\mathrm{tr}\sqrt{\gamma^{-1}\omega}\Big)^3 \notag \\
                                            &&-3\Big(L/N+\mathrm{tr}\sqrt{\gamma^{-1}\omega}\Big)\Big(L^2/N^2+\mathrm{tr}\big(\gamma^{-1}\omega \big)\Big)+2\Big(L^3/N^3+\mathrm{tr}\big(\gamma^{-1}\omega \big)^{3/2}\Big) \Big\} \notag \\
                                            &=&\sqrt{\det \gamma}\, \Big\{\frac{1}{2}L\Big({\mathrm{tr}}^2\sqrt{\gamma^{-1}\omega}-\mathrm{tr}\big(\gamma^{-1}\omega \big)\Big) \notag \\
                                            &&+\frac{1}{6}N\Big({\mathrm{tr}}^3\sqrt{\gamma^{-1}\omega}-3\mathrm{tr}\sqrt{\gamma^{-1}\omega}\, \mathrm{tr}\big(\gamma^{-1}\omega \big)+2\mathrm{tr}\big(\gamma^{-1}\omega \big)^{3/2}\Big)\Big\} \ .
\end{eqnarray}
This is also linear with respect to the lapse functions.
The last one is 
\begin{eqnarray}
\sqrt{-\det g}\, e_4\big(\sqrt{g^{-1}f}\big)=\sqrt{-\det g}\,\det{\sqrt{g^{-1}f}}=\sqrt{-\det f}=L\, \sqrt{\det \omega}  \ .
\end{eqnarray}

To sum up, the interaction terms read
\begin{align}
&\sum_{n=0}^{4}\beta_n e_n\big(\sqrt{g^{-1}f}\big) \notag \\
=&N\, \sqrt{\det \gamma}\Big[\beta_0+\beta_1\mathrm{tr}\sqrt{\gamma^{-1}\omega}+\frac{1}{2}\beta_2\Big({\mathrm{tr}}^2\sqrt{\gamma^{-1}\omega}-\mathrm{tr}\big(\gamma^{-1}\omega \big)\Big) \notag \\
&\qquad \qquad \qquad \frac{1}{6}\beta_3\Big({\mathrm{tr}}^3\sqrt{\gamma^{-1}\omega}-3\mathrm{tr}\sqrt{\gamma^{-1}\omega}\, \mathrm{tr}\big(\gamma^{-1}\omega \big)+2\mathrm{tr}\big(\gamma^{-1}\omega \big)^{3/2}\Big)\Big] \notag \\
+&L\, \Big[\sqrt{\det \gamma}\Big\{\beta_1+\beta_2\mathrm{tr}\sqrt{\gamma^{-1}\omega}+\frac{1}{2}\beta_3\Big({\mathrm{tr}}^2\sqrt{\gamma^{-1}\omega}-\mathrm{tr}\big(\gamma^{-1}\omega \big)\Big)\Big\}+\beta_4 \sqrt{\det \omega}\Big] \ .
\end{align}
Notice that all interaction terms are linear in $L$ and $N$. 
This is the advantage of mini-superspace model, which makes the Hamailtonian constraint analysis simple. 

In the paper, for simplicity, we always assume  that interactions are minimal~\cite{Hassan2011a, Nima2012}, namely
 \begin{align}
 \beta_0=3,\quad   \beta_1=-1,\quad   \beta_2=0,\quad  \beta_3=0, \quad \beta_4=1   \ .
 \end{align}
Clearly, this simplification does not lose any generality concerning with the ghost analysis.

\section{Constraint Algebra}

In this appendix, we calculate a Poisson bracket.
It is sufficient to look at the following 
\begin{align}
\big\{C_N\, ,\, C_L\big\} =&\Big\{\frac{M_g^2}{\sqrt{\det \gamma}}\Big(\frac{1}{2}{\pi^i}_i{\pi^j}_j-\pi^{ij}\pi_{ij}\Big)\, ,\, -a_1\sqrt{\det \gamma}\Big\} \notag \\
                           &+\Big\{-a_1\sqrt{\det \gamma}\, \mathrm{tr}\sqrt{\gamma^{-1}\omega}\, ,\, \frac{M_f^2}{\sqrt{\det \omega}}\Big(\frac{1}{2}{p^i}_i{p^j}_j-p^{ij}p_{ij}\Big)\Big\} \notag \\
                           =&a_1\bigg(\frac{M_g^2}{\sqrt{\det \gamma}}\Big\{\sqrt{\det \gamma}\, ,\, \frac{1}{2}{\pi^i}_i{\pi^j}_j-\pi^{ij}\pi_{ij}\Big\}
                           -M_f^2\sqrt{\frac{\det \gamma}{\det \omega}}\, \Big\{\mathrm{tr}\sqrt{\gamma^{-1}\omega}\, ,\, \frac{1}{2}{p^i}_i{p^j}_j-p^{ij}p_{ij}\Big\}\bigg)  \ .
\end{align}
The point is that the result is proportional to $a_1$.
Each term can be manipulated as
\begin{eqnarray}
\Big\{\sqrt{\det \gamma}\, ,\, \frac{1}{2}{\pi^i}_i{\pi^j}_j-\pi^{ij}\pi_{ij}\Big\}&=&\frac{\partial \sqrt{\det \gamma}}{\partial \gamma_{mn}}\, \frac{\partial}{\partial \pi^{mn}}\Big(\frac{1}{2}{\pi^i}_i{\pi^j}_j-\pi^{ij}\pi_{ij}\Big) \notag \\
                                                                                   &=&\frac{1}{2}\sqrt{\det \gamma}\, \gamma^{mn}\, \big(\gamma_{mn}{\pi^i}_i-2\pi_{mn}\Big) \notag \\
                                                                                   &=&\frac{1}{2}\sqrt{\det \gamma}\, {\pi^i}_i 
\end{eqnarray}
and
\begin{eqnarray}
\Big\{\mathrm{tr}\sqrt{\gamma^{-1}\omega}\, ,\, \frac{1}{2}{p^i}_i{p^j}_j-p^{ij}p_{ij}\Big\}
&=&\frac{\partial \mathrm{tr}\sqrt{\gamma^{-1}\omega}}{\partial \omega_{mn}}\, \frac{\partial}{\partial p^{mn}}\Big(\frac{1}{2}{p^i}_i{p^j}_j-p^{ij}p_{ij}\Big) \notag \\
&=&\frac{1}{2}\big(\sqrt{\gamma^{-1}\omega}^{-1}\gamma^{-1}\big)^{mn}\, \big(\omega_{mn}{p^i}_i-2p_{mn}\big) \notag \\
&=&\frac{1}{2}{p^i}_i\, \mathrm{tr}\sqrt{\gamma^{-1}\omega}-\mathrm{tr}\big(\sqrt{\gamma^{-1}\omega}\, p\, \omega\big)  \ ,
\end{eqnarray}
where $p$ represents a matrix with components $p^{mn}$. 

In this case, $C_{NL} \neq 0$ because there is an interaction between $g$ and $f$, namely $a_1 \neq 0$.
Thus, if the Poisson bracket is non-trivial or not is determined by the interaction pattern. 


\providecommand{\noopsort}[1]{}\providecommand{\singleletter}[1]{#1}%

\end{document}